# The Flexible Plasma Trap (FPT) for the production of overdense plasmas

S. Gammino,[a] L. Celona,[a] D. Mascali,[a] G. Castro,[a,1] G. Torrisi,[a] L. Neri,[a] M. Mazzaglia,[a,b] G. Sorbello,[a,c] O. Leonardi,[a] L. Allegra,[a] G. Calabrese,[a] F. Chines,[a] G. Gallo[a] and S. Passerello[a]

[a] *INFN-LNS,*
*Via S. Sofia 62, 95123, Catania, Italy*

[b] *Universitá degli studi di Catania, Dipartimento di Fisica e Astronomia,*
*Via S. Sofia 64, 95123 Catania, Italy*

[c] *Universitá degli studi di Catania, Dip. di Ingegneria Elettrica Elettronica e Informatica,*
*Viale A. Doria 6, 95125, Catania, Italy*

E-mail: castrog@lns.infn.it

Abstract: Electron Cyclotron Resonance Ion Sources are currently the most efficient ion sources among those used in facilities dedicated to nuclear physics. The need for a more flexible magnetic field and RF injection system suggested to design and develop a different type of plasma trap, named Flexible Plasma Trap (FPT). The magnetic field of FPT is generated by means of three coils while microwaves in the range 4–7 GHz can be injected by three different inputs, one placed along the axis and two placed radially. FPT can work in different plasma heating schemes so it will be an ideal tool for studies of plasma and multidisciplinary physics. Moreover, a microwave launcher has been designed and installed to the FPT for launching microwaves with a variable tilt angle with respect to the magnetic field. This paper describes the characteristics of the FPT along with the preliminary results of plasma diagnostics.

Keywords: Ion sources (positive ions, negative ions, electron cyclotron resonance (ECR), electron beam (EBIS)); Plasma diagnostics — charged-particle spectroscopy; Plasma diagnostics — interferometry, spectroscopy and imaging; Plasma diagnostics - probes

---

[1]Corresponding author.

# Contents

## 1 Introduction

The Electron Cyclotron Resonance Ion Sources (ECRIS) are used to generate high intensity highly charged ion beams characterized by high stability and high reliability. The improvement of ECRIS performances followed up to now the so called Geller's scaling laws: both extracted current $I_{\text{extr.}}$ and mean charge state $<q>$ depend on the microwaves frequency as [1]:

$$I_{\text{extr.}} \propto \omega^2 \tag{1.1}$$

$$<q> \propto \log\left(\omega^{\frac{7}{2}}\right) \tag{1.2}$$

Additional limitations are given by the high B-mode (HBM) concept [2] that sets a rule of thumb for the value of confining magnetic field, above $2B_{\text{ECR}}$. Any further step forward in terms of extracted current and charge states needs a further increase of microwaves frequency and magnetic field strength, involving a considerable impact on the design complexity and cost. Moreover, from a technological point of view, the required superconducting magnets are close to the limit of current technology.

The use of new schemes of plasma heating, and in particular the excitation and absorption of Electron Bernstein Waves (EBW) [3], could be an alternative to overcome the present limitations. Preliminary tests, carried out with a fixed field plasma reactor, have shown that it is possible to increase by a factor 5 to 10 the electron density above the density cut-off [4, 5]. The tests have shown that the tuning of the magnetic field profile and of microwaves frequency plays a key role in boosting the plasma density and temperature.

The experience gained over the last years led to the development of Flexible Plasma Trap (FPT), characterized by a huge flexibility in terms of magnetic field configuration, microwaves frequency and possibility to launch microwaves with different frequency, propagation directions and polarization. The tuning of the magnetic field profile and microwaves' frequency makes possible the generation of plasmas characterized by different density, temperature and ion lifetime, thereby optimizing the production of high currents of protons, light ions or $H_2^+$ and $H_3^+$ beams.

The large range of electron density and temperature of the generated plasmas makes FPT the proper set-up for studies of astrophysical, nuclear physics and multidisciplinary physics. In last years, in fact, the interest of the scientific community for studies of fundamental physics in a plasma environment has been growing. The plasma is not only a source of ions to be accelerated, but also a fertile environment for studies of fundamental and applied physics. In this perspective, a special attention has been paid to the diagnostics issues, to make possible the full characterization of the electron energy distribution function in all energetic domains of the electron and ion populations.

## 2   Alternative mechanisms of plasma heating

An option to overcome the limits of the ECR heating consists in the use of plasma waves, having no cutoffs in a plasma environment. A plasma wave is a rarefaction-compression wave whose electric field is parallel to the propagation direction. Typically, a large number of plasma oscillation modes can be excited in a plasma; among them, the EBW are the most promising because they are strongly absorbed by the plasma at cyclotron harmonics [6]. Due to their electrostatic nature, EBWs must be generated inside the plasma from electromagnetic (EM) waves. In particular, an extraordinary wave (usually named as X wave) converts in an EBW at the upper hybrid resonance.

There exist three different conversion mechanisms to convert EM waves to EBWs: "*the high field side launching conversion*", the "*Fast X- EBW (FX-B) conversion*" and "*the O- Slow X- EBW (O-SX-B) conversion*" [7].

- *High field side launching* (arrow 1 in figure 1): X waves are launched by regions where $B/B_{\text{ECR}} > 1$. X waves are here not screened by the R cut-off, reach the UHR crossing the ECR from the high field side, then being converted into EBW. Because of the characteristics of the magnetic field of FPT (see section 3.1), this mechanism can not be applied to our device.

- *Fast-X-B conversion* (arrow 2 in figure 1): the fast X-mode (FX) tunnels through the evanescent region between the R-wave cut-off and the UHR and couples to the slow X-mode (SX) that, in turn, converts to EBWs at UHR.

    Direct FX-B conversion heating is used in experiments with relatively low magnetic field, where the normalized gradient length $k_0 L_n$ is ~ 0.3 ($k_0$ being the wave number of the incident wave in vacuum ad $L_n = n_e/(\partial n_e/\partial x)$ the length-scale of the electron density) [7, 8].

- *O-Slow X-B conversion* (arrow 3 in figure 1): the R cut-off is crossed by the O wave that, if the conditions for O-SX conversion are valid at the O cut-off, is converted into SX waves which are in turns converted into Bernstein waves at UHR.

    The efficiency of O-SX transition process is maximized for $k_0 L_n \sim$ 1–20, i.e. for flattened density profiles [7, 9].

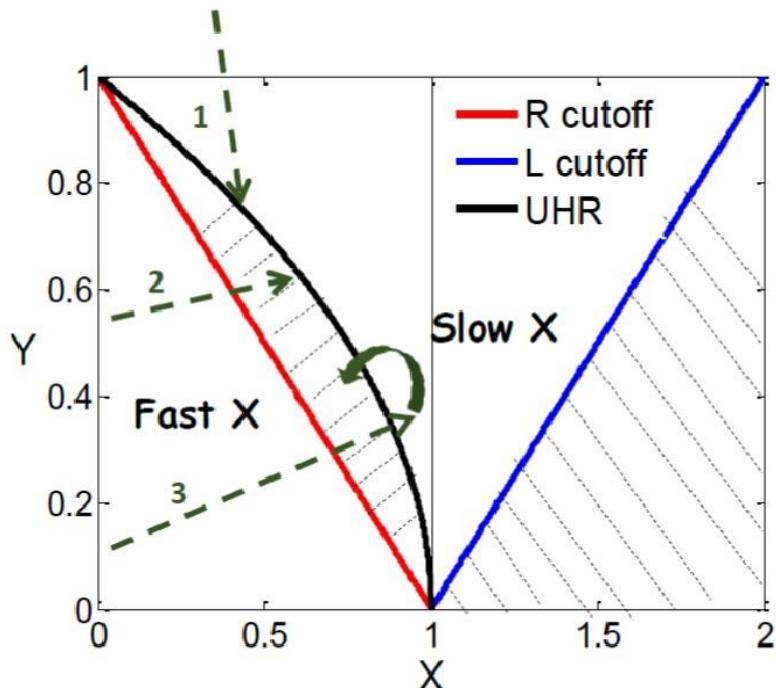

**Figure 1**. Main conversion mechanisms from X wave to EBW.

As it will be described in the next sections, the flexibility of the FPT magnetic field, as well as of the RF launching systems, will allow to explore both XB or OXB conversion mechanisms. The first results from ongoing experiments are expected within 2017.

## 3 The flexible plasma trap

The FPT has been successfully installed at INFN-LNS and the commissioning phase is almost complete. In this section the main characteristics of the FPT are highlighted.

### 3.1 The magnetic field

The FPT magnetic field is generated by means of three solenoids which allow the tuning of the magnetic field profile. The solenoids have been developed in order to allow the generation of the following three magnetic field profiles:

- The *Off-Resonance configuration* (dashed-dot line in figure 2) permits to study the plasma dynamics in a configuration like that of Microwave Discharge Ion Source (MDIS) [10]. The typical shape of the magnetic field for MDIS machines is a quasi-flat profile everywhere above the resonance value. This ensures electron densities around the density cut-off and temperatures sufficient for hydrogen ionization ($T_e \sim$ 15–20 eV). Such plasma parameters require RF power around 0.5 to 1.5 kW and background pressures down to $10^{-5}$–$10^{-4}$ mbar. The main advantage stays in the high stability of the extracted beam and in its low emittance.

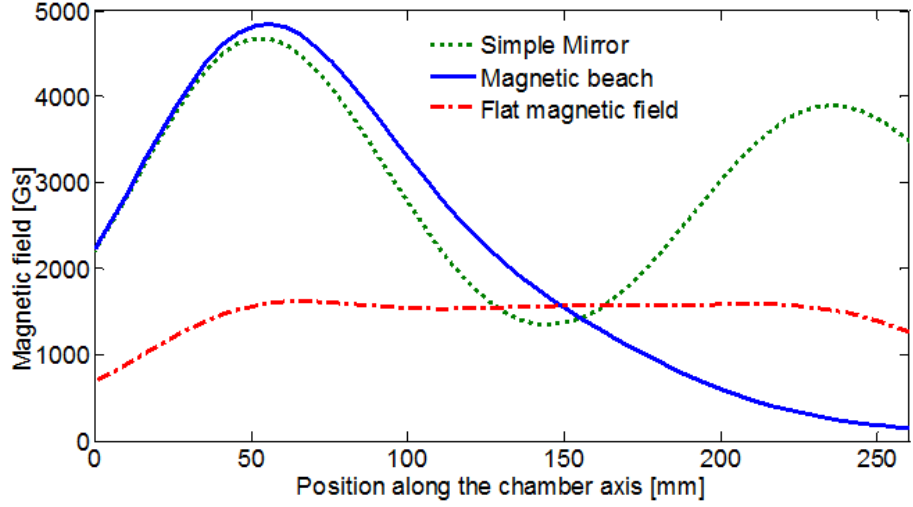

**Figure 2**. Off-resonance, simple mirror and magnetic beach field profiles generated by the three coils of FPT.

- The *Simple mirror configuration* (dotted line in figure 2) permits to increase the ion lifetime for multiply charged ions production. Ion lifetime $\tau_i$, in fact, depends on the ratio between $B_{\max}$ and $B_{\min}$ as:

$$\tau_i \propto \log\left(\frac{B_{\max}}{B_{\min}}\right) \quad (3.1)$$

Studies about balance equations of the different species within plasma for proton beams production ($H_3^+$, $H_2^+$, $H^+$) or for multicharged ions from light elements like Li or C, reveal that the reciprocal abundance is regulated by the relative lifetimes. In a quasi-flat magnetic field, under normal operational pressure conditions, ions lifetime is only governed by collisional diffusion across the magnetic field, which is a rather fast process. The increase of ion lifetime, obtained when using the simple-mirror configuration, will increase the ionization efficiency thus allowing the production of $2^+$ or $3^+$ ions already at moderate RF power. Simple mirror configuration enables also to generate an almost flat plasma density that can be used to excite an overdense plasma via O-SX-B conversion.

- The *Magnetic Beach configuration* (dashed line in figure 2) enables to study the mechanisms of electromagnetic to electrostatic conversion. The electromagnetic waves requires a rapidly dropping magnetic field and a high density gradient which makes possible either upper hybrid resonance and second harmonic absorption [7] thus making possible the conversion from electromagnetic wave to Bernstein wave via the FX-B conversion.

## 3.2 The plasma chamber

A particular attention was paid to the design of the plasma chamber. It gives the chance to couple three different waveguide inputs, each one perpendicular to the other (figure 3) and to have an adequate water cooling in the microwave windows location permitting at the same time to host different type of diagnostics.

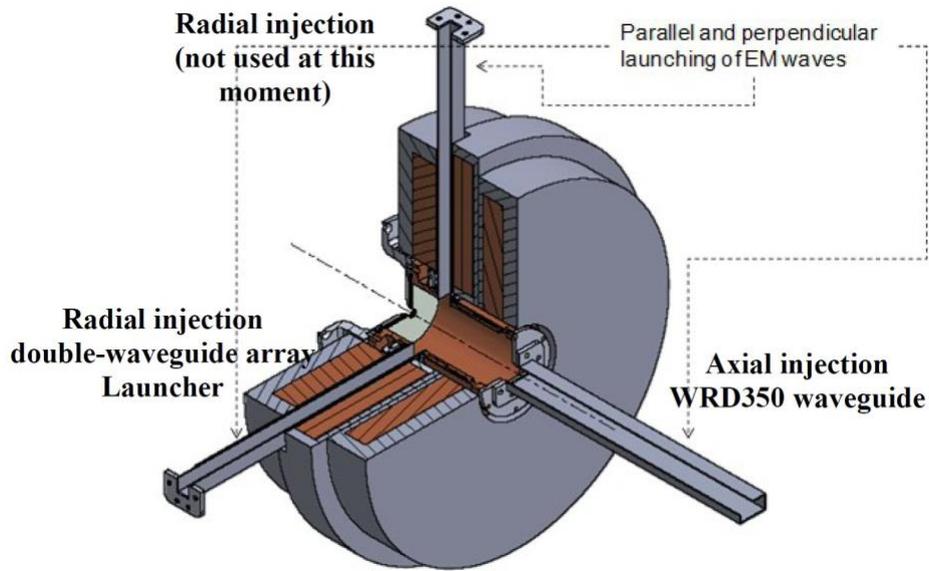

**Figure 3**. Plasma chamber and microwave inputs in FPT.

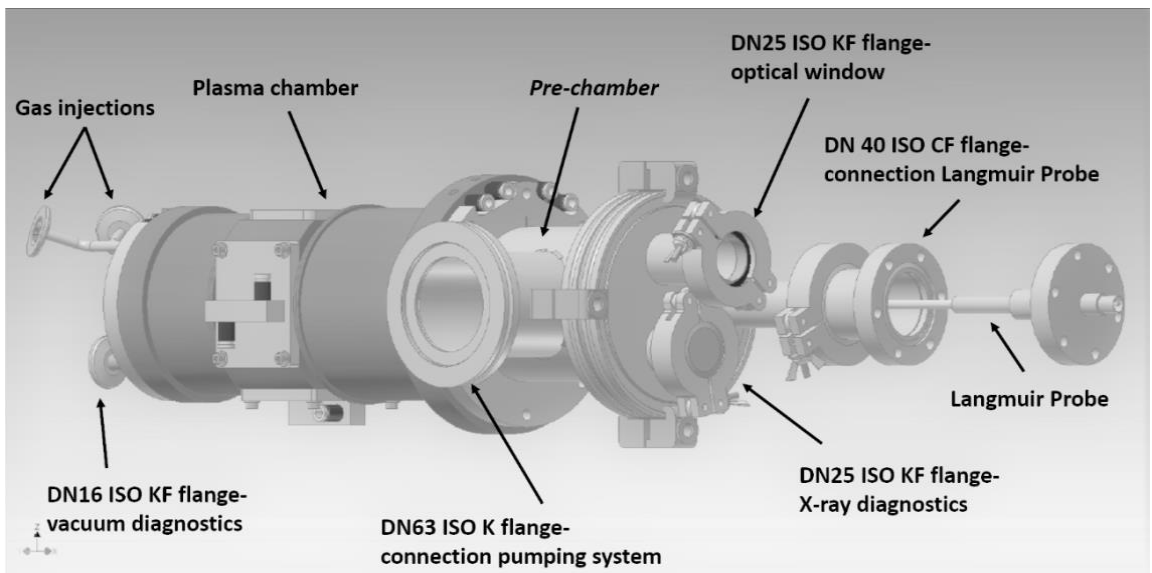

**Figure 4**. Render view of the FPT plasma chamber.

The plasma chamber is shown in figure 4. Its dimensions are 82.0 mm in diameter and 260.1 mm in length. It is made of oxygen free high conductivity copper and the tolerances of the different pieces are dictated by the needs of microwave to plasma coupling, in any case better than 0.1 mm.

The plasma chamber has been equipped by a stainless-steel made "pre-chamber" in order to host the vacuum system and the diagnostics tools. In this way, the vacuum pumping is performed through a 60 mm DN flange placed drilled in the lateral walls of the pre-chamber, while the three DN40 flanges placed on the endplate are free to host the diagnostic tools that will be described in section 4.

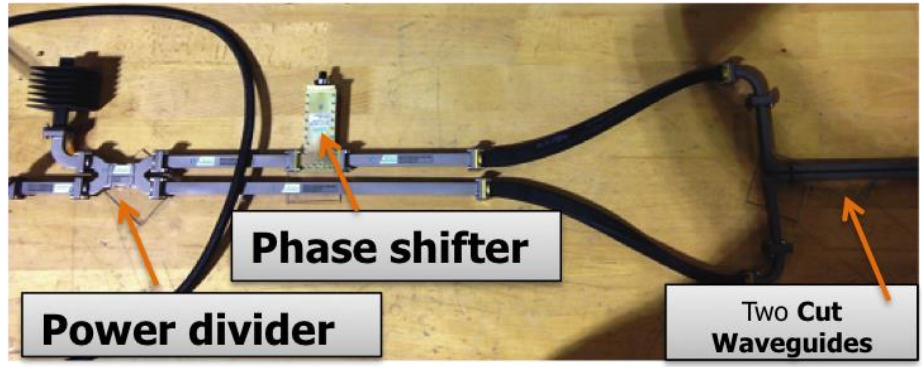

**Figure 5**. The microwave launcher showing the power divider, phase shifter, and two-cut waveguide array [11].

### 3.3 The injection system

The FPT plasma can be generated by means of three different microwave systems, through a parallel and two perpendicular microwaves injection. The axial injection is operated at frequencies ranging in the range of 4–7 GHz. The signal is generated by a *Rohde & Schwarz* generator, amplified by a TWT and sent to the FPT by WRD350 waveguides.

A directional coupler enables the measure of forward and reflected power, while an insulator safeguards the TWT by the power reflected backwards.

The perpendicular microwaves' injection, working at 14 GHz frequency, consists of an Anritsu generator and a CPI klystron and the launcher shown in figure 5. (The other perpendicular microwave injection is not used at this moment)

The use of two different microwave frequencies enables exploring also the aspects of resonant absorption of waves' energy by electrons at the second-harmonic of the cyclotron resonance.

The contemporary use of the two microwaves' input allows operating with a double mode: a fraction of power provided to the plasma by means of the usual ECR-heating process by a standard double-ridge rectangular waveguide, and the remaining amount of power injected along the perpendicular direction with respect to the axis-symmetric magnetic field, in order to excite EBW through O-SX-B conversion mechanism.

In reference [7], H. Laqua shows that the injection angle plays a key role for modal conversion from the Ordinary to SX wave. The perspective of modifying the injection angle in FPT led us to the development and construction of an innovative launcher for the injection along the perpendicular direction [11].

It consists of an array of two properly phased rectangular WR62 with their small side parallel to the magnetic field direction of the plasma chamber of FPT. The two waveguides are driven in $TE_{10}$ mode with relative phase controlled by a calibrated phase shifter, a loaded four-port power divider, and two flexible and twistable WR62 waveguide complete the launcher layout. The antenna has been fixed inside the transversal microwave injection camera port. The microwaves' transmission line is shown in figure 5.

The antenna pattern was investigated both theoretically and experimentally. The experimental characterization is in good agreement with the numerical simulations (see figure 6) and it has shown that, setting up a suitable phase difference, it is possible to tilt the angle of maximum radiation up to 48° in order to direct the microwaves towards the different areas of the plasma chamber.

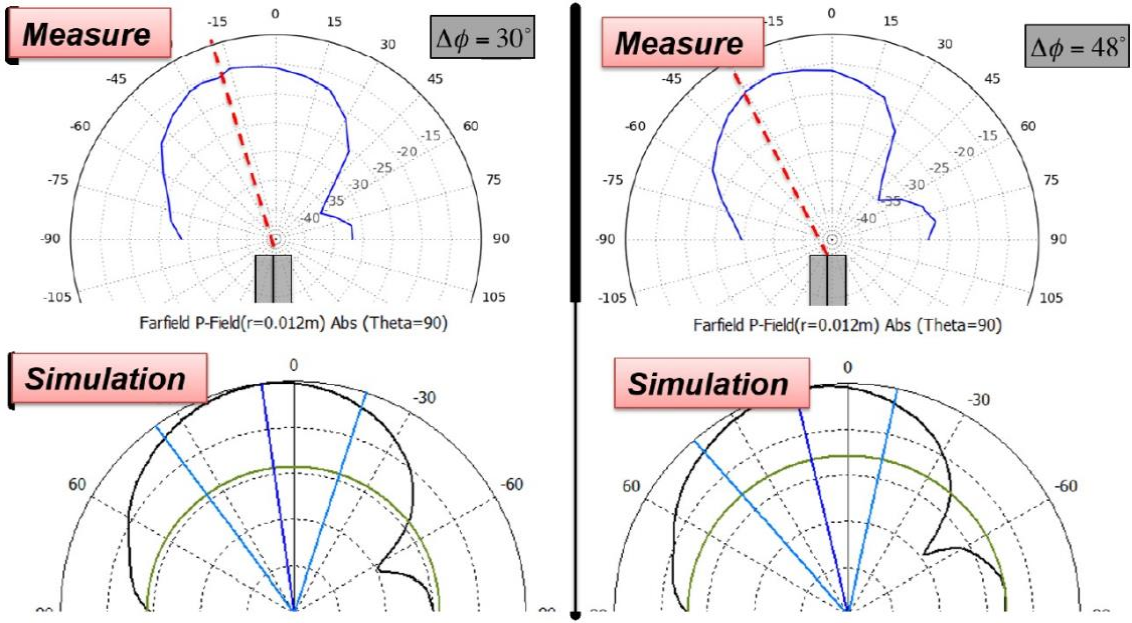

**Figure 6**. Experimental measurements vs simulation for a phase shift Δϕ = 30° and Δϕ = 48°.

## 4 The diagnostics

A special attention has been paid to diagnostics issues. This is a critical point for developing future highly performing ion sources. Several ion beam features directly follow from plasma parameters, and they can be optimized by modifying "on-line" the plasma properties, if such a properties are known with great accuracy.

FPT diagnostics includes both invasive and non-invasive diagnostics. The goals of the authors, however, is to develop non-invasive diagnostics in order to avoid, in a few years, any use of invasive diagnostics, which could affect the results of the measurement or be damaged in presence of high density plasmas, in case of EBW plasmas generations.

Hereinafter, the diagnostics tools already implemented in FPT:

- *Langmuir probe (LP) diagnostics* allows obtaining electron density and temperature measurements of bulk electrons (1–100 eV) from the numerical analysis of the resistivity curve.

  Data are numerically analysed by means of different LP models validated in different density ranges of interest [4]. LP diagnostics feature some limitations: the probe undergoes plasma damage in high density, high temperature plasmas ($n_e > 5 \cdot 10^{17}$ m$^{-3}$, $t_e > 100$ eV) as the ones expected in case of EBW heating. Furthermore, the LP perturbs the plasma chamber, especially by the electromagnetic point of view and the values of electron density and temperature depend strongly by the physical model used to determine them from resistivity curve. Because of these limits, further diagnostics have been developed.

- *Volume-integrated X-ray spectroscopy* in low energy domain (2–30 keV, by using SDD detectors) or highly energetic regimes (> 30 keV, by using HpGe detectors). It allows obtaining electron temperature and density in the energy domain of the warm and hot electron population [12];

- A *pin-hole camera* is being used for the direct detection of the spatially-resolved spectral distribution of X-rays produced by the electronic motion. This diagnostics has been already applied to ECRIS plasmas [13]. This diagnostics is particularly suitable in case of EBW generation, since it permits to evaluate the plasma regions where microwaves are coupled to plasmas, and indeed to identify the resonance regions (see also reference [4]) .

- *Optical emission spectroscopy* has been developed in order to characterize the extremely low energy part (< 10 eV) of the electron energy distribution function. It allows obtaining not only the electron density and temperature of cold electrons in the plasma bulk, but also the relative percentage of the ion species within the plasma [14].

- Microwave interferometry enables measuring the overall density of the whole plasma. The frequency-sweep interferometer, named VESPRI, has been already tested in the plasma reactor [15] and allowed the measurement of overdense plasma density generated by EBW by means of the so-called frequency-sweep method [5].

  Two conical-horn antennas launch into the plasma chamber a probing signal in the range 18.5–26.5 GHz. The signal is synthesized by a frequency-sweep oscillator. The superposition of the reference and plasma leg signals produces a beating signal, whose Fourier transform gives the information about electron density [5]

Figure 7 shows a render view of the FPT during the commissioning phase with three diagnostics connected to the "diagnostics flange", in particular the optical spectrometer, the X-ray detector and the LP are visible in the picture.

## 5 Preliminary results

During the commissioning phase, we started to characterize FPT in simple mirror and B-flat configuration, modifying source parameters as microwave power and frequency, neutral pressure and magnetic configuration.

Characterization started by using LP and non-invasive diagnostics. However we plan to avoid the LP use soon, and characterize plasma only by means of non-invasive tools.

Figure 8 shows the electron density of an Argon plasma, calculated by means of OML model [4] in simple mirror configuration and 6.8 GHz microwaves frequency, when increasing microwave power from 10 to 60 W (neutral pressure $8 \cdot 10^{-5}$ mbar). Plasma is well-confined within the ECR layer and density does not overcome the density cut-off in any point ($n_{cutoff} = 5.8 \cdot 10^{17}$ m$^{-3}$).

Preliminary results from X-ray diagnostics are labelled in figure 9 and show the amount of X radiation collected by the SDD detector versus microwaves' power in the range 1–30 keV. X-rays have been collimated by means of a 1 mm aluminium collimator, in order to make possible the measurement of the solid angle covered by the detector.

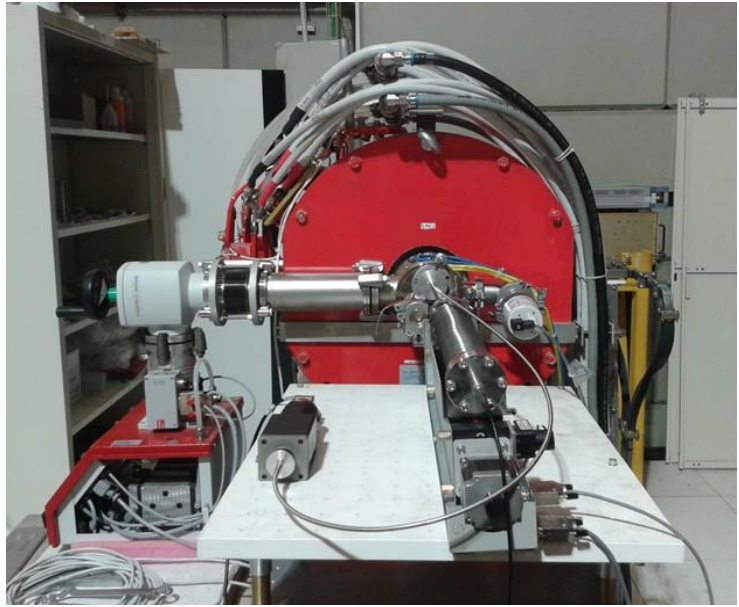

**Figure 7**. A view of the FPT and three diagnostics system.

The rate of X rays increase strongly with microwaves' power, putting in evidence the increasing of the higher energy electron component.

Interesting information has been carried out by means of the optical spectrometry. Figure 10 shows the spectrum of a hydrogen plasma when modifying the magnetic field configuration. The ratio $B_{max}/B_{min}$ has been changed from 1.14 (almost flat magnetic field) to 1.48 (simple mirror configuration) in order to study the influence of the mirror ratio on the ion and neutral species ($H$, $H_2$, $H^+$ and $H_2^+$) composing the plasma. The spectrum has been analysed by means of the radiative collisional model, which allows to relate the intensity of Balmer lines and Fulcher band to the relative abundances of the atomic versus molecular hydrogen $n_H/n_{H_2}$ [14]. In particular, $n_H/n_{H_2}$ can be calculated by the ratio between the Balmer γ line and the integral of Fulcher band in the range 600–650 nm. Also, information about electron density and temperature can be obtained by the ratio between Balmer α and β lines or Balmer β and γ lines. A detailed description of the line ratio method for hydrogen can be found in reference [15] Figure 10 shows a typical spectrum obtained in FPT with the over mentioned diagnostics. Balmer lines and Fulcher band are highlighted. Figure 11 shows the preliminary results in different source conditions. Figure 11a shows the dependence of $n_H/n_{H_2}$ on the mirror ratio $B_{max}/B_{min}$ of the magnetic field, when FPT is operated in simple mirror configuration. As $B_{max}/B_{min}$ increases, $n_H/n_{H_2}$ increases too. This results can be explained by the increase of confinement times $\tau_{conf.}$ due to the increase of the mirror ratio ($\tau_{conf.} \propto B_{max}/B_{min}$). Figure 11b show the dependence of $n_H/n_{H_2}$ on the neutral pressure in simple mirror configuration. As expected, larger values of $H_2$ pressure within FPT corresponds to lower values of $n_H/n_{H_2}$. The experimental OES campaign is currently in progress and new results will be published soon.

The microwave launcher has been already tested and characterized in empty chamber conditions [11]. Preliminary tests for coupling the 14 GHz microwaves by launcher to plasma by means

– 9 –

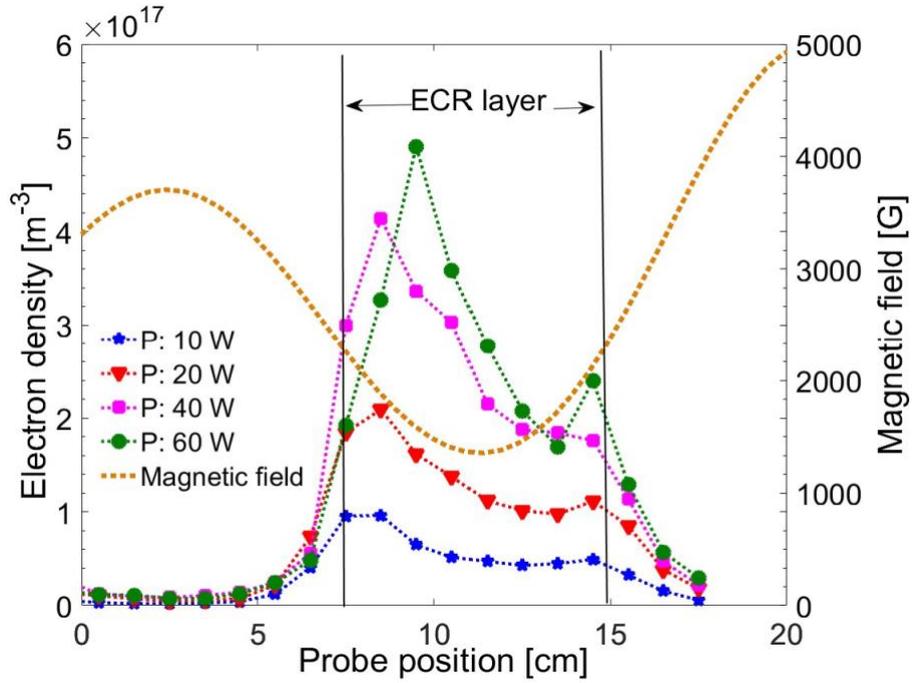

**Figure 8**. Electron density with respect to the probe position for different microwave powers in simple mirror configuration.

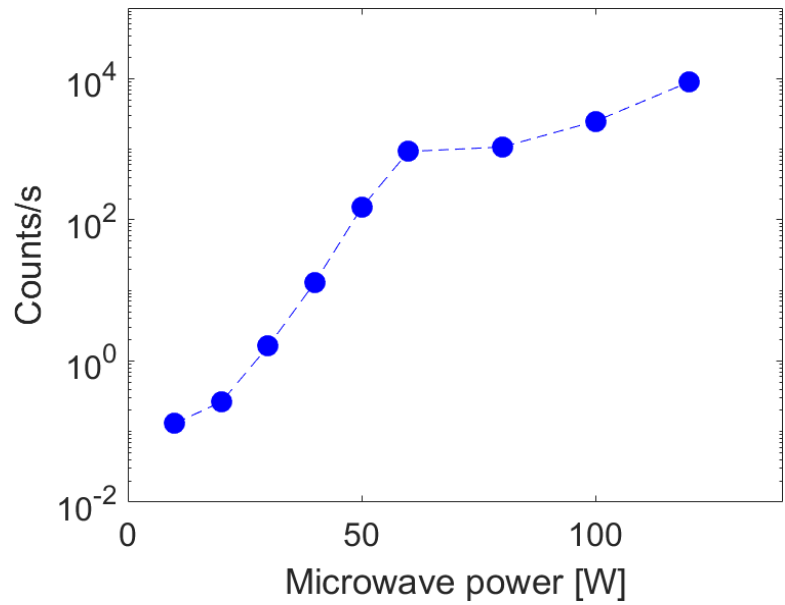

**Figure 9**. Overall X ray emission in the range 1–30 keV, in simple mirror configuration (6.8 GHz frequency, $8 \cdot 10^{-5}$ mbar pressure) when increasing microwave power up to 120 W.

of OXB modal conversion are being carried out in these months.

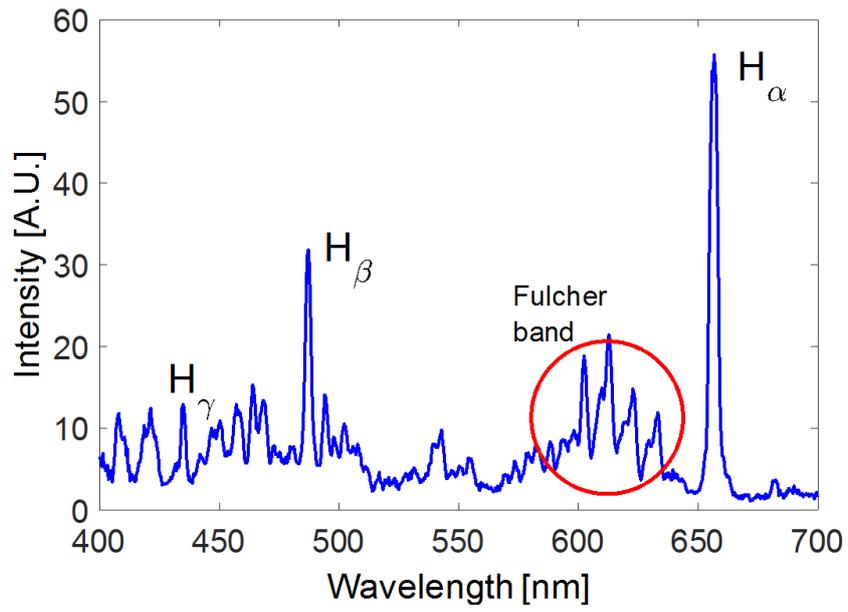

**Figure 10**. Hydrogen spectrum obtained for $B_{max}/B_{min}$ = 1.2. The Balmer lines and the Fulcher band are highlighted.

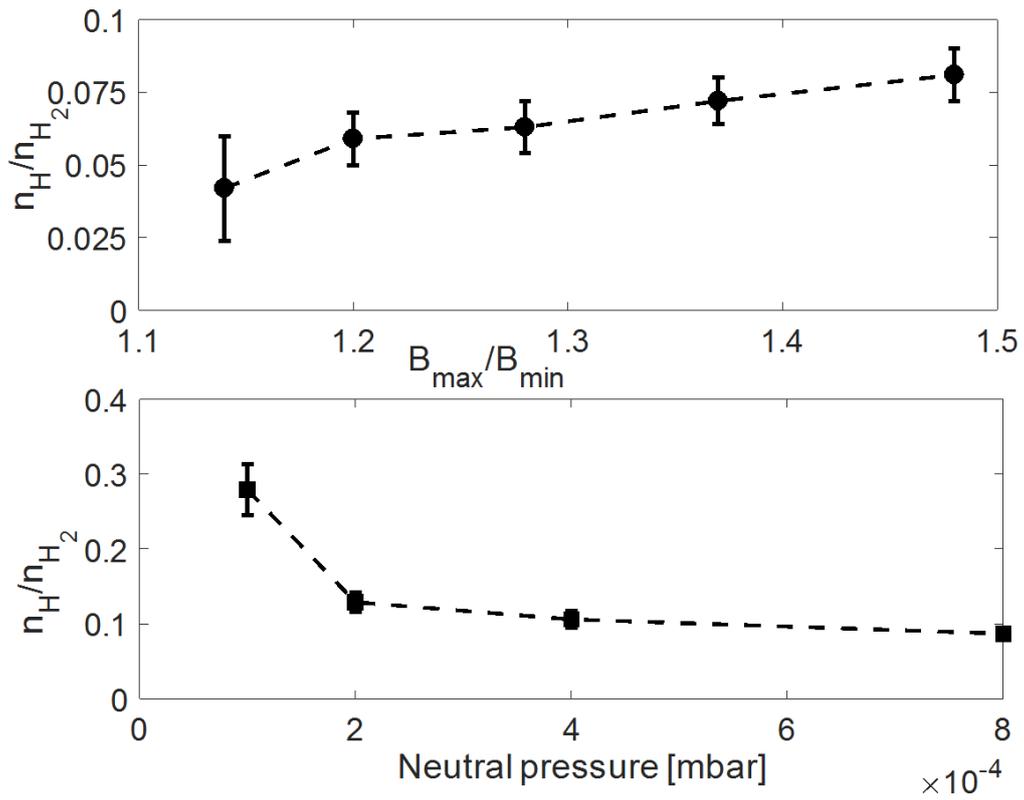

**Figure 11**. A) $n_H/n_{H_2}$ ratio versus the ratio $B_{max}/B_{min}$. B) $n_H/n_{H_2}$ ratio versus the neutral pressure.

# 6 Conclusions

A multi-purpose plasma trap has been designed and constructed at INFN-LNS. The commissioning phase confirmed the possibility to generate a plasma in different magnetic configurations and microwave frequencies.

Its flexibility permits the deeper study of the mechanisms for overdense plasmas generation and to investigate new strategies for improving the plasma-wave coupling. A novel microwave launcher has been installed for tilting the microwave launching angle within FPT in order to permit the OXB conversion and indeed the generation of overdense plasmas by EBW heating. The installation of different diagnostics permits to investigate the electron energy distribution function in different energy domains, from a few eV (O.E.S. and Langmuir probe diagnostics) to tens or hundreds of keV (X-ray diagnostics, pin-hole cameras), while the interferometric technique allows to characterize the whole plasma population along a sight line.

In perspective, FPT represents a "full-optional" test bench for the development of innovative plasma heating schemes, and testing of new kinds of diagnostics to be exported to microwave discharge ion source and ECRIS.

The use of different types of diagnostics, furthermore shall improve our knowledge of plasma heating and help to further develop the techniques of plasma heating already tested in the plasma reactor, aimed to the generation of higher brightness beams and multiply charged ions build up in absence of confining radial magnetic field.

## Acknowledgments

The support of the 5[th] National Committee of INFN through the RDH and UTOPIA experiments is gratefully acknowledged. The support of G. Manno, S. Marletta, A. Maugeri, A. Seminara, G. Pastore and S. Vinciguerra is warmly acknowledged.

## References

[1] R. Geller, *Electron Cyclotron Ion sources and ECR plasmas*, IOP Publishing Ltd, (1996).

[2] S. Gammino and G. Ciavola, *The role of microwave frequency on the high charge states build-up in the ECR ion sources*, *Plasma Source Sci. Technol.* **5** (1996) 19.

[3] I.B. Bernstein, *Waves in a Plasma in a Magnetic Field*, *Phys. Rev. Lett.* **109** (1958) 10.

[4] G. Castro, *Overdense plasma generation in a compact ion source*, *Plasma Sources Sci. Technol.* **26** (2017) 055019.

[5] D. Mascali et al., *The first measurement of plasma density in an ECRIS-like device by means of a frequency-sweep microwave interferometer*, *Rev. Sci. Instrum.* **87** (2016) 095109.

[6] Taylor and J.S. Wills, *A high-current low-emittance dc ECR proton source*, *Nucl. Instrum. Meth.* **A 309** (1991) 37.

[7] H.P. Laqua, *Electron Bernstein wave heating and diagnostic*, *Plasma Phys. Control. Fusion* **49** (2007) R1.

[8] G. Torrisi, G. Sorbello, O. Leonardi, D. Mascali, L. Celona and S. Gammino, *A new launching scheme for ECR plasma based on two-waveguides-array*, *Microw. Opt. Tech. Lett.* **58** (2016) 2629.

[9] K.S. Golovanivsky, V.D. Dougar-Jabon and D.V. Reznikov, *Proposed phisical model for very hot electron shell structures in electron cyclotron resonance-driven plasmas*, *Phys. Rev.* **E 52** (1995) 2969.

[10] A.K. Ram and S.D. Schultz, *Excitation, propagation, and damping of electron Bernstein waves in tokamaks*, *Phys. Plasmas* **7** (2000) 4084.

[11] Y.Y. Podoba et al., *Direct Observation of Electron-BernsteinWave Heating by O-X-B-Mode Conversion at Low Magnetic Field in the WEGA Stellarator*, *Phys. Rev. Lett.* **98** (2007) 255003.

[12] D. Mascali et al., *X-ray spectroscopy of warm and hot electron components in the CAPRICE source plasma at EIS testbench at GSI*, *Rev. Sci. Instrum.* **85** (2014) 02A956.

[13] D. Mascali et al., *Electron cyclotron resonance ion source plasma characterization by X-ray spectroscopy and X-ray imaging*, *Rev. Sci. Instrum.* **87** (2016) 02A510.

[14] U. Fantz et al., *Spectroscopy — a powerful diagnostic tool in source development*, *Nucl. Fusion* **46** (2006) S297.

[15] G. Torrisi et al., *Microwave frequency sweep interferometer for plasma density measurements in ECR ion sources: Design and preliminary results*, *Rev. Sci. Instrum.* **87** (2016) 02B909.